\documentclass[%
 aip,
 amsmath,amssymb,
 reprint,%
]{revtex4-1}

\usepackage{graphicx}
\usepackage{dcolumn}
\usepackage{bm}

\usepackage[utf8]{inputenc}
\usepackage[T1]{fontenc}
\usepackage{mathptmx}
\usepackage{etoolbox}

\usepackage{amsmath}
\usepackage{mathrsfs}
\usepackage{amssymb}
\usepackage{amsfonts}
\usepackage{graphicx}
\usepackage{float}
\usepackage{graphicx}
\usepackage{subfigure}
\usepackage{epstopdf}

\makeatletter
\def\@email#1#2{%
 \endgroup
 \patchcmd{\titleblock@produce}
  {\frontmatter@RRAPformat}
  {\frontmatter@RRAPformat{\produce@RRAP{*#1\href{mailto:#2}{#2}}}\frontmatter@RRAPformat}
  {}{}
}%
\makeatother
\begin{document}

\preprint{AIP/123-QED}

\title{Second-order Cumulants Ghost Imaging}
\author{Huan Zhao}
\affiliation{ 
Department of Physics, Changchun University of Science and Technology,Changchun 130022, P. R. China
}%

\author{Xiao-Qian Wang}%
 \altaffiliation{ Corresponding author: xqwang21@163.com. }
 
\affiliation{ 
Department of Physics, Changchun University of Science and Technology,Changchun 130022, P. R. China
}%
\author{Chao Gao}
\affiliation{ 
Department of Physics, Changchun University of Science and Technology,Changchun 130022, P. R. China
}%
\author{Zhuo Yu}%

\affiliation{ 
Department of Physics, Changchun University of Science and Technology,Changchun 130022, P. R. China
}%
\affiliation{ 
College of Physics and Electronic Information, Baicheng Normal University, Baicheng 137000 P. R. China
}%
\author{Shuang Wang}%
 
\affiliation{ 
Department of Physics, Changchun University of Science and Technology,Changchun 130022, P. R. China
}%
\author{Li-Dan Gou}%
 
\affiliation{ 
Department of Physics, Changchun University of Science and Technology,Changchun 130022, P. R. China
}%
\author{Zhi-Hai Yao}
 \homepage{Corresponding author: yaozh@cust.edu.cn.}
\affiliation{ 
Department of Physics, Changchun University of Science and Technology,Changchun 130022, P. R. China
}%

\date{\today}

\begin{abstract}
In the conventional ghost imaging (GI), the image is retrieved by correlating the reference intensity fluctuation at a charge-coupled device (CCD) with the signal intensity fluctuation at a bucket detector. In this letter, we present the protocol of GI, it is called Second-order Cumulants ghost imaging (SCGI). The image is retrieved by the fluctuation information of correlating intensity fluctuation at two detectors, and the resolution limit can be enhanced than conventional GI. The experimental results of SCGI agreement with theoretical results.
\end{abstract}

\maketitle
Ghost imaging (GI) is an imaging technique based on correlation measurement\cite{pittman1995optical,gao2019ghost,gao2017optimization,gatti2004ghost,ribeiro1994controlling,gatti2003entangled,meyers2008ghost,cheng2004incoherent,abouraddy2001role,baleine2006correlated}. The GI experiment has been demonstrated by Pittman et al\cite{pittman1995optical}. Due to its surprising nonlocal feature, it was once considered as an unique phenomenon of quantum entanglement\cite{abouraddy2001role}. Later, Bennink et al. presented theoretical arguments and provided experimental demonstration that GI can be performed with a classical source\cite{bennink2002two}. Compared with conventional imaging technique, GI has many advantages such as high spatial resolution\cite{gong2012experimental},  high robustness\cite{cheng2009ghost,meyers2011turbulence} and so on.\par
The classical limit of resolution of an optical instrument was formulated in the well-known works by Abbe and Rayleigh. This classical limit states that the resolution of an optical system is limited by diffraction on the system pupil. The resolution is determined by the point spread functions (PSF) of the system. The narrower the PSF, the better the resolution\cite{chen2017sub,wang2019enhancement}. In GI system, the resolution is superior to the classical Rayleigh limit\cite{gong2012experimental}.
The reason is that more nature of the light source can be distinguished by second-order coherence function than that by first-order coherence function\cite{scully1999quantum}, that is, second-order coherence properties of the light source has more information than the first-order coherence properties. Many schemes of resolution enhancement proposals have been suggested, such as compressive sensing technique\cite{gong2012experimental,du2012influence}, non-Rayleigh speckle field\cite{kuplicki2016high}, low-pass spatial filter scheme\cite{chen2017sub,meng2018super}, high-pass spatial-frequency field scheme\cite{sprigg2016super} and so on.
The cumulants of measured intensity distributions is applied to traditional imaging system and it has been proposed to narrow the PSF curve by Li et al. in 2019\cite{li2019beyond}.
The result shows that such cumulants can beat the classical Rayleigh limit. In this letter, second-order cumulants of Hanbury Brown and Twiss (HBT) intensity correlation measurement is applied to reconstruct the image of the object. The result shows that the resolution limit of GI can be enhanced by this protocol. The optical setup of GI is not changed by our scheme.
\par
\begin{figure}[ht!]
\centering
\includegraphics[width=8 cm]{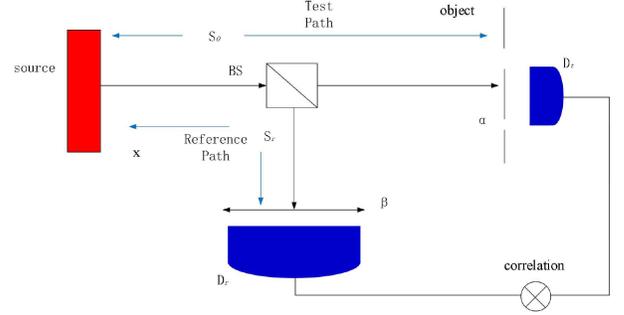}
\caption{Experimental setup of the GI. An illumination light is split at a 50:50 beam splitter (BS) into two paths. In the test path, the light illuminates the object and then is collected by a bucket detector with no spatial resolution. In the reference path, the light intensity distribution is recorded by a charge-coupled device (CCD) camera.}\label{qvh}
\label{fig:false-color}
\end{figure}
A typical GI experimental setup is shown in Fig.~\ref{qvh}. In general, the double slit is used for measured the object, when the resolution has been studied by many researchers\cite{zeng2017influence,ferri2005high,zhang2009improving}. 
In this system, the source is monochromatic light with a wavelength $\lambda$. A light beam in the test path propagates to the object through an optical system with a PSF
\begin{equation}\label{(1)}
h_{t}(x,\alpha)=\frac{e^{-iks_{o}}}{i\lambda s_{o}}\exp(\frac{-i\pi}{\lambda s_{o}}(x-\alpha)^{2}).
\end{equation}
Here $k=\frac{2\pi}{\lambda}$, $s_{o}$ is the distance of the test path of optical system, $x$ and $\alpha$ are the transverse coordinates on the source and the object plane, respectively. If $T(\alpha)$ represents the transmission function of the object, the light field behind the object is given by
\begin{equation}\label{(2)}
E(\alpha)=\int\ E(x)h_{t}(x,\alpha)T(\alpha)\,dx,
\end{equation}
where $E(x)$ denotes the light field of source plane at $x$. The optical-electric current operator at the bucket detector is
\begin{equation}\label{(3)}
B_{t}=\int\ E^{*}(\alpha)E(\alpha)\,d\alpha.
\end{equation}
Suppose that the PSF of the reference path is $h_{r}(x,\beta)$ which has similar form as Eq.~(\ref{(1)}). The light field on the CCD plane is given by
\begin{equation}\label{(4)}
E(\beta)=\int\ E(x)h_{r}(x,\beta)\,dx,
\end{equation}
where $\beta$ is the transverse coordinates on the CCD plane. Thus, the optical-electric current operator at the CCD plane is
\begin{equation}\label{(5)}
I(\beta)=E^{*}(\beta)E(\beta).
\end{equation}
The correlation between the intensity fluctuations at the reference and test detectors is
\begin{align}\label{(6)}
\Delta G^{(2)}(\beta)& =\langle [I(\beta)-\langle I(\beta)\rangle][B_{t}-\langle B_{t}\rangle]\rangle =\langle \Delta I(\beta)\Delta B_{t}\rangle \notag \\
=& \int|\int\ G^{(1)}(x,x^{'})T(\alpha)h_{t}(x,\alpha)h^{*}_{r}(x^{'},\beta)\,dxdx^{'}|^{2}d\alpha \notag \\
&-\langle I(\beta)\rangle\langle B_{t}\rangle,
\end{align}
where $\langle \dots\rangle$ represents the ensemble average. $G^{(1)}(x,x^{'})=\langle E^{*}(x)E(x^{'})\rangle$ is the first-order correlation function at source. We consider the case where the illuminated light is a point-like source which randomly and uniformly distributed on the source plane. If the light spot is located at $x_{0}$, we may have $G^{(1)}(x,x_{o})=I_{0}\delta (x-x_{0})$, where $I_{0}$ is the intensity of source. Here we also consider the distances from source to the object and the source to the CCD are same, namely $s_{r}=s_{o}=z$, we can get
\begin{equation}\label{(7)}
\Delta G^{(2)}(\beta)=I_{0}^{2} \int\ |T(\alpha)|^{2}\text{sinc}^{2}(\frac{2\pi R}{\lambda z} (\alpha-\beta))\,d\alpha.
\end{equation}
Where $R$ is the radius of the source plane. Obviously, the image resolution is confined by this PSF. From Eq.~(\ref{(7)}), it is $\text{sinc}^{2}(\frac{2\pi R}{\lambda z}(\alpha-\beta))$ that produces the Airy disk and the first zero of $\text{sinc}^{2}$ function leads to the resolution limit.\par
In general, $I_{0}$ is taken as a constant in the coventional GI under the assumption of the emitting power of the light source is perfect stable. In fact, the emitting power of the light source can not keep stable.
Thus, the $\Delta G^{(2)}(\beta)$ in Eq.~(\ref{(7)}) should be substituted with $\Delta G^{(2)}(I_{0},\beta)$. The fluctuation of $I_{0}$ lead to the fluctuation of $\Delta G^{(2)}(I_{0}, \beta)$. We use the concept ‘Cumulants’, which is always used in statistics, to describe the fluctuation of $\Delta G^{(2)}(I_{0}, \beta)$ which have more information than $\Delta G^{(2)}(I_{0}, \beta)$. The narrower PSF can be got by the cumulants than by  $\Delta G^{(2)}(I_{0}, \beta)$. So the resolution limit of GI can be improved. The cumulant-generating function $K(s, \beta)$ is defined.
\begin{equation}\label{(8)}
K(s, \beta)=\ln (\langle \exp(s\Delta G^{(2)}(I_{0},\beta))\rangle).
\end{equation}
The nth-order cumulants is given by
\begin{equation}\label{(9)}
\kappa_{n}(\beta)=\frac{d^{(n)}K(s,\beta)}{ds^{(n)}}|_{s=0}.
\end{equation}
We consider second-order cumulants, which can be written as
\begin{align}\label{(10)}
 &\kappa_{2}(\beta)=\langle [\Delta G^{(2)}(I_{0}, \beta)-\langle \Delta G^{(2)}(I_{0}, \beta)\rangle ]^{2}\rangle \notag\\
 &=\int \kappa_{2}(\alpha,\beta)\,d\alpha +L(\beta),
\end{align}
where
\begin{align}\label{(11)}
 \kappa_{2}(\alpha,\beta)& =\langle[\Delta G^{(2)}(I_{0},\alpha,\beta)-\langle\Delta G^{(2)}(I_{0},\alpha,\beta)\rangle]^{2}\rangle \notag \\
 &=\langle [I_{0}^{2}-\langle I_{0}^{2}\rangle]^{2}\rangle\times |T(\alpha)|^{4}\times \notag \\
 &\text{sinc}^{4}(\frac{2\pi R}{\lambda z}(\alpha-\beta)),
\end{align}
\begin{align}\label{(12)}
L(\beta) &= \int_{\alpha}\int_{\alpha^{'}\neq \alpha}\ \langle [\Delta G^{(2)}(I_{0},\alpha,\beta)-\langle\Delta G^{(2)}(I_{0},\alpha,\beta)\rangle]\notag\\
[\Delta & G^{(2)}(I_{0},\alpha ^{'},\beta) -\langle\Delta G^{(2)}(I_{0},\alpha ^{'},\beta)\rangle]\rangle \,d\alpha d\alpha^{'} \notag\\
=\langle &[I_{0}^{2}-\langle I_{0}^{2}\rangle]^{2}\rangle\times \int_{\alpha}\int_{\alpha^{'}\neq\alpha}|T(\alpha)|^{2}|T(\alpha^{'})|^{2}\times  \notag\\
\text{sinc}&^{2}(\frac{2\pi R}{\lambda z}(\alpha-\beta))\text{sinc}^{2}(\frac{2\pi R}{\lambda z}(\alpha ^{'}-\beta))\,d\alpha d\alpha^{'}.
\end{align}
Where $\Delta G^{(2)}(I_{0},\alpha,\beta)=I_{0}^{2} |T(\alpha)|^{2}\text{sinc}^{2}(\frac{2\pi R}{\lambda z} (\alpha-\beta))$ which is the correlation between the intensity fluctuations at $\alpha$ and $\beta$.
We find that $\kappa_{2}(\beta)$ is the fluctuation information of $\Delta G^{(2)}(I_{0}, \beta)$, and $\kappa_{2}(I_{0},\alpha,\beta)$ is the fluctuation information of $\Delta G^{(2)}(I_{0},\alpha,\beta)$. $L(\beta)$ is the cross-information generated by total of correlating $\Delta G^{(2)}(I_{0},\alpha,\beta)-\langle \Delta G^{(2)}(I_{0},\alpha,\beta) \rangle$ with $\Delta G^{(2)}(I_{0},\alpha^{'},\beta)-\langle \Delta G^{(2)}(I_{0},\alpha^{'},\beta) \rangle$ for all different $\alpha$ and $\alpha^{'}$ $(\alpha^{'}\neq\alpha)$. We use $\kappa_{2}(\beta)$ instead of $\Delta G^{(2)}(I_{0}, \beta)$ to reconstruct the image of the object. The scheme is called Second-order Cumulants ghot imaging (SCGI). \par
\begin{figure}[h]
  \centering
  \includegraphics[width=7 cm,height=5 cm]{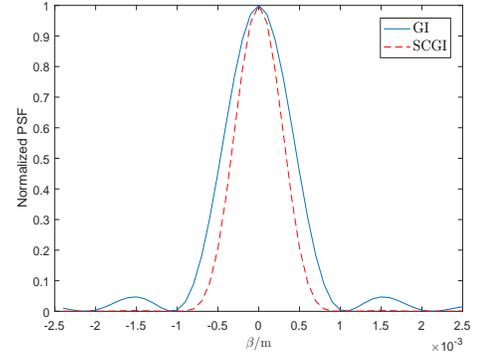}\\
  \caption{When $\lambda = 532nm, z=5m$ and $R=0.0025m$, the PSF of a pinhole-like object at $\alpha=0$ by GI (the blue solid curve) and SCGI (the red dash curve), respectively;}. \label{2}
  \label{fig:false-color}
\end{figure}

From Eqs.~(\ref{(10)}-\ref{(12)}), we can get the intensity PSF of SCGI as $\text{sinc}^{4}$ function. As for conventional GI, the form of PSF is $\text{sinc}^{2}$ function. The full-width at half-maximum (FWHM) of PSF in Eq.~(\ref{(7)}) is wider than that in Eq.~(\ref{(10)}). The results are shown in Fig.~\ref{2}.\par
\begin{figure*}
  \centering
  \subfigure[]{\includegraphics[width=6cm]{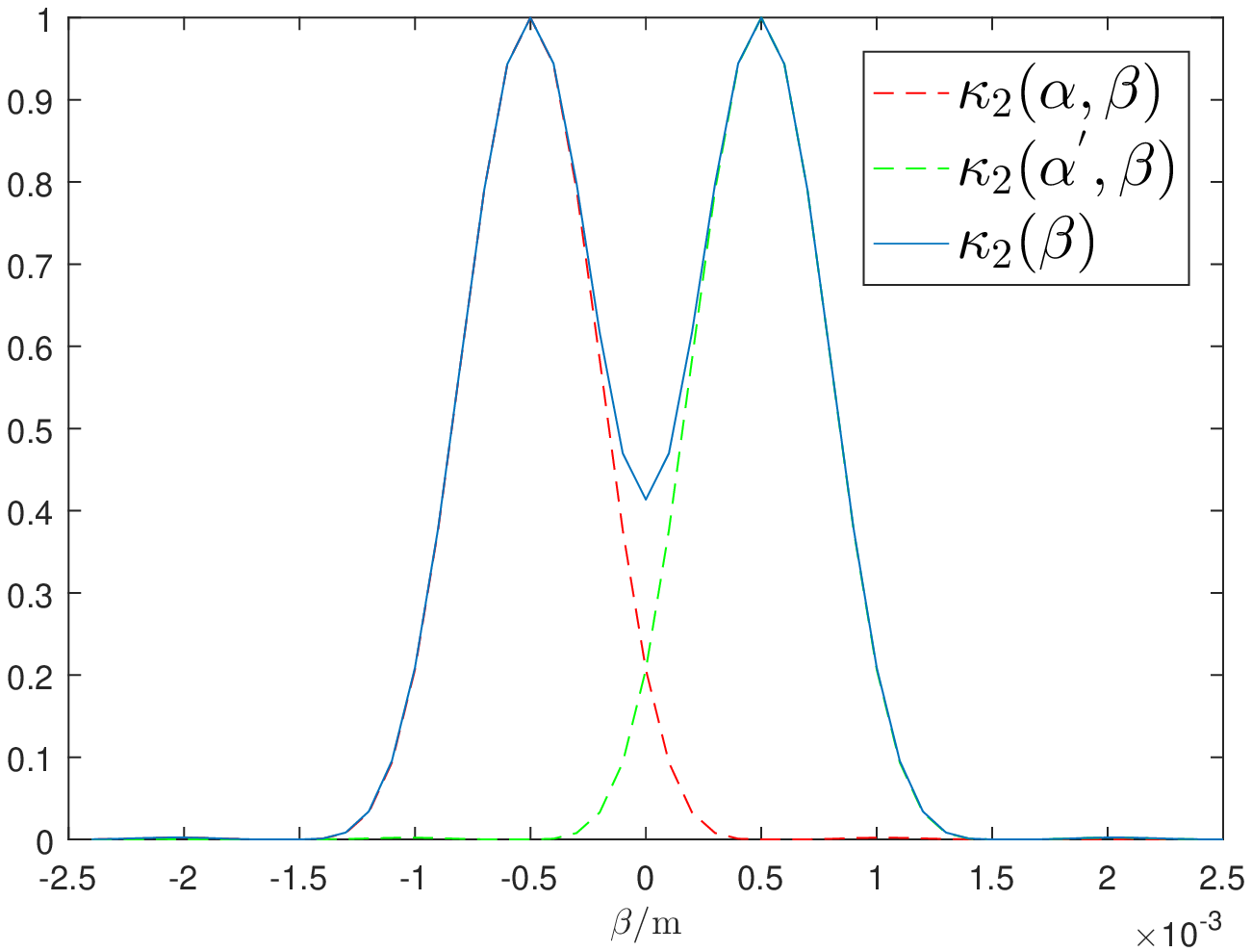}}
  \hspace{0in}
  \subfigure[]{\includegraphics[width=6cm]{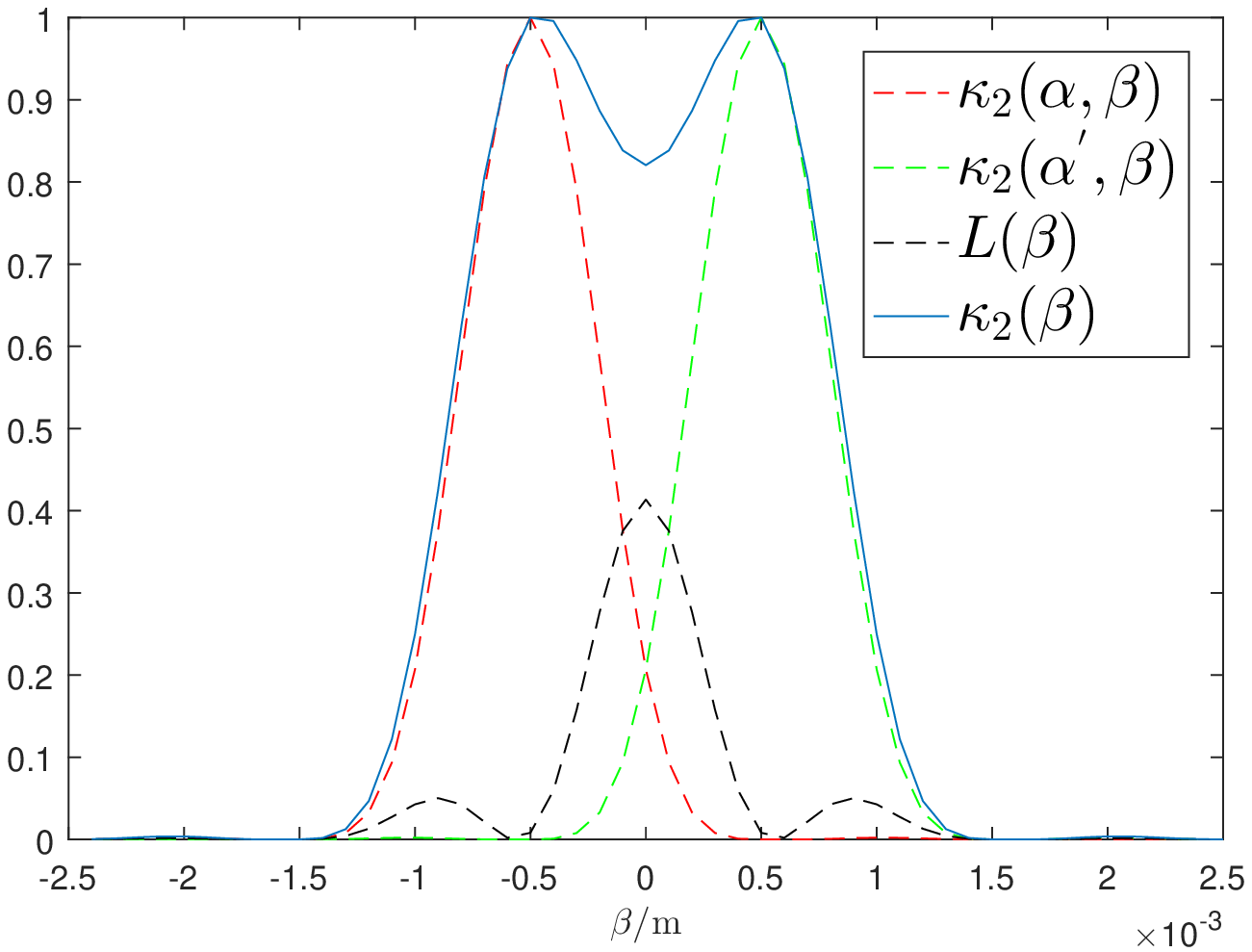}}
  \centering
  \caption{The image of two-pinholes object by SCGI when $\alpha=-0.0005m$, $\alpha ^{'}=0.0005m$, $\lambda=532nm$, $R=0.0025m$ and $z=5m$; (a)$L(\beta)=0$; (b)$L(\beta)\neq 0$;}\label{3}
  \label{fig:false-color}
\end{figure*}

From Eq.~(\ref{(10)}), we find that $L(\beta)$ affects the resolution of SCGI. In order to understand how $L(\beta)$ affects the resolution, we study the case where the object is two-pinholes. The two pinholes are at $\alpha$ and $\alpha^{'}=-\alpha$, respectively. The image of the object as shown in Fig.~\ref{3}. From Fig.~\ref{3}(a) and Fig.~\ref{3}(b), we can find the  image of the object when $L(\beta)=0$ is clear than it when $L(\beta) \neq 0$. This is because the cross-information which cannot be distinguished between $\kappa_{2}(\alpha, \beta)$ and $\kappa_{2}(\alpha^{'}, \beta)$ is generated when $\langle [\Delta G^{(2)}(I_{0},\alpha, \beta)-\langle \Delta G^{(2)}(I_{0},\alpha, \beta)\rangle][\Delta G^{(2)}(I_{0},\alpha^{'}, \beta)-\langle \Delta G^{(2)}(I_{0},\alpha^{'}, \beta)\rangle]\rangle \neq 0$. \par
\begin{figure*}[ht!]
\centering
  \subfigure[]{\includegraphics[width=5cm]{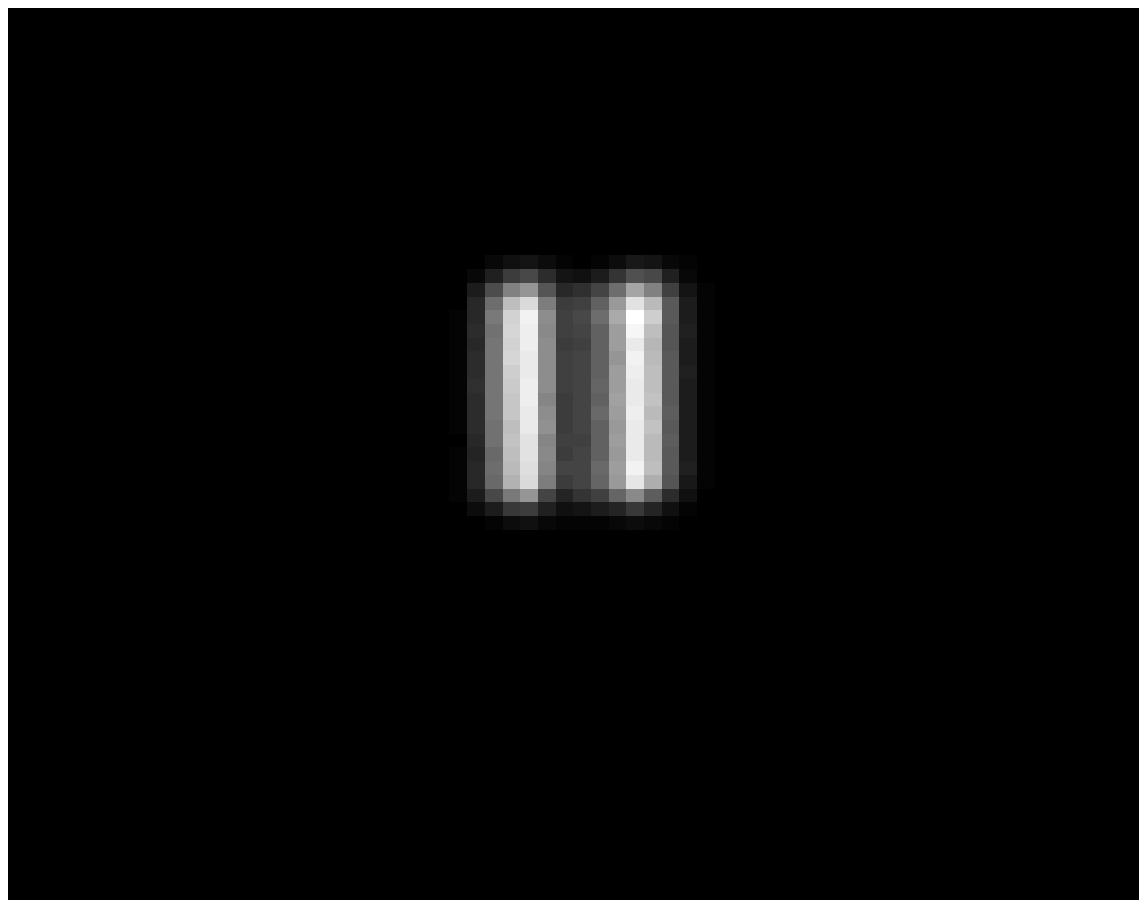}}
  \hspace{0in}
 \subfigure[]{\includegraphics[width=5cm]{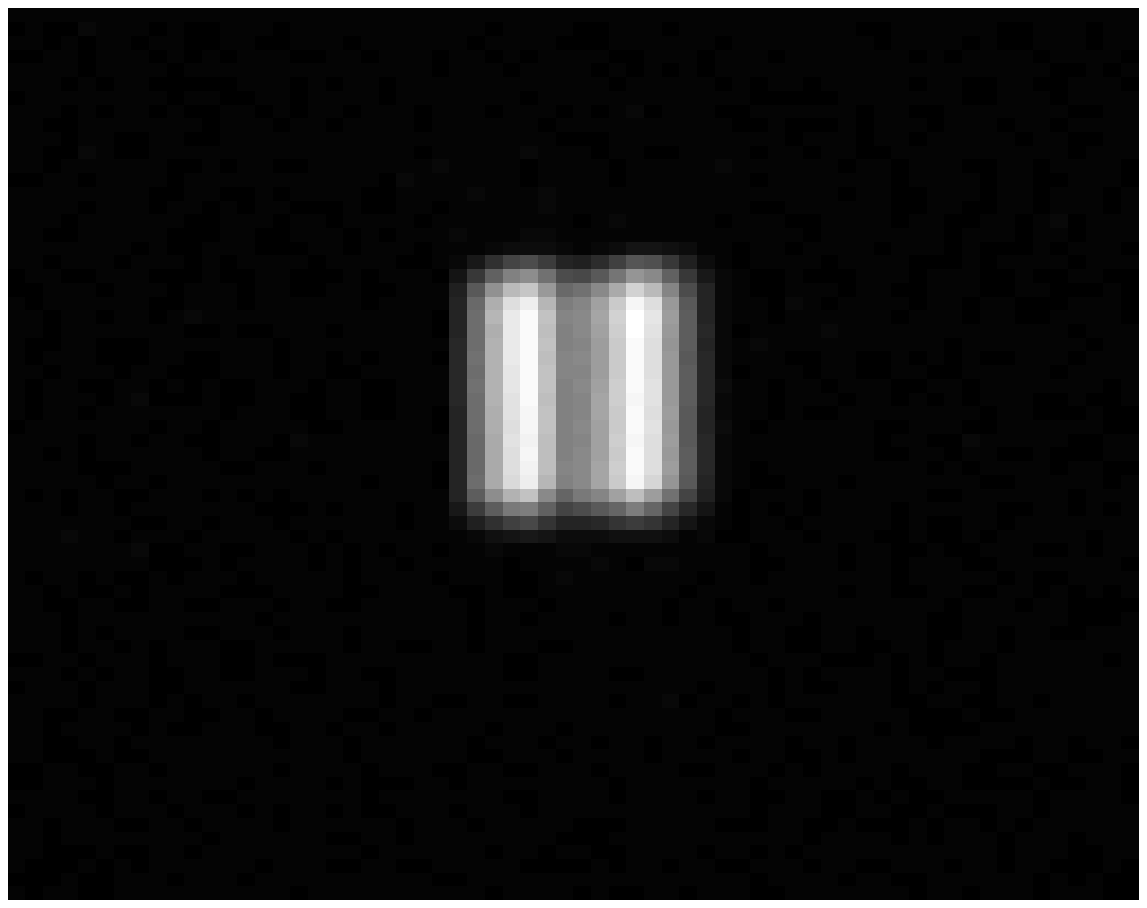}}
  \hspace{0in}
  \subfigure[]{\includegraphics[width=5cm]{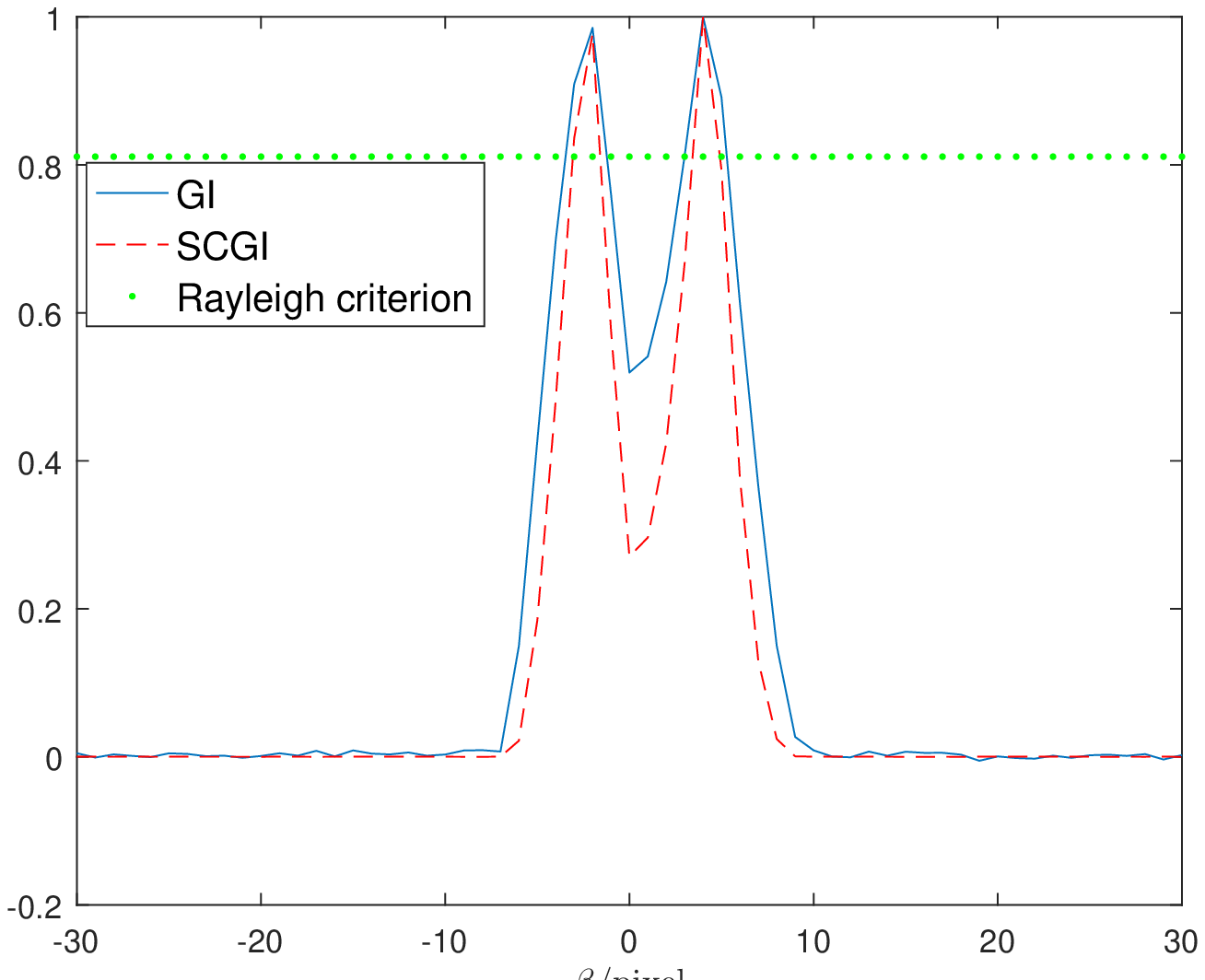}}
 \hspace{0in}
 \subfigure[]{\includegraphics[width=5cm]{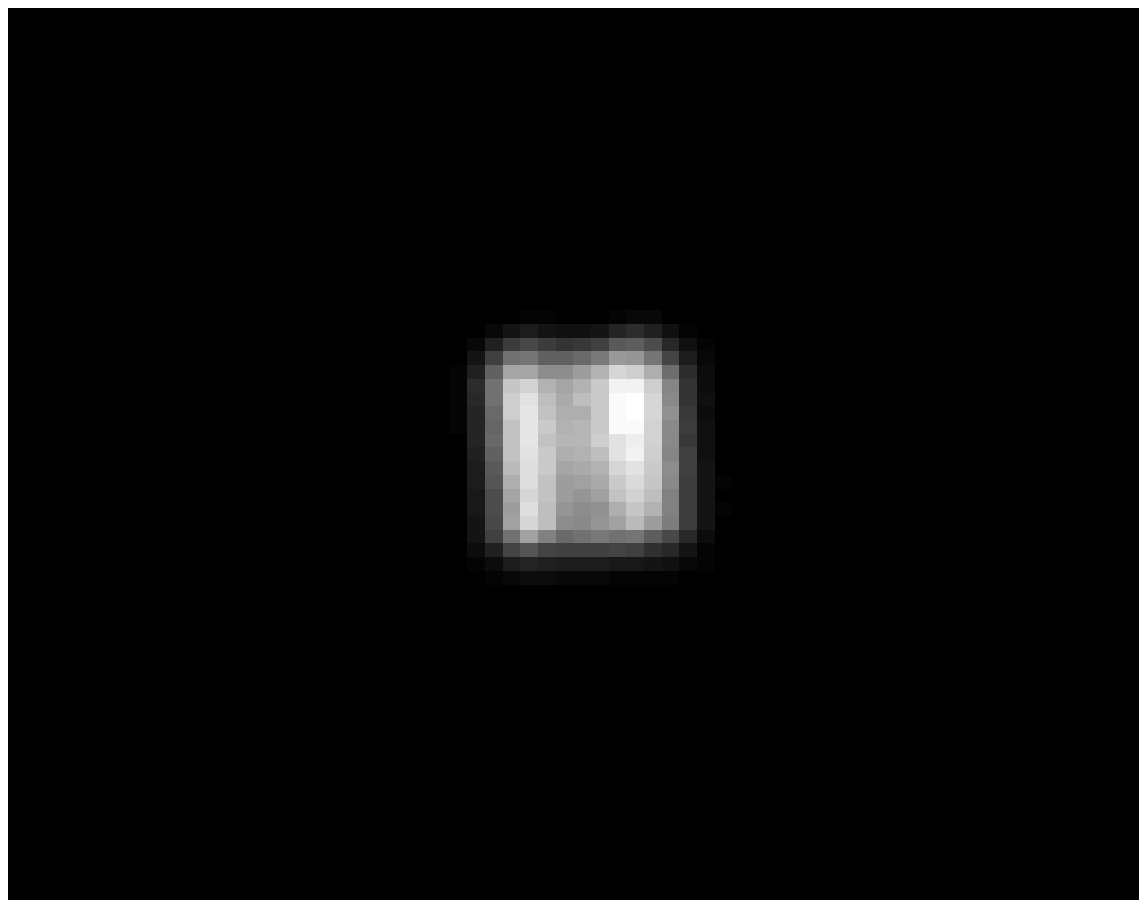}}
  \hspace{0in}
  \subfigure[]{\includegraphics[width=5cm]{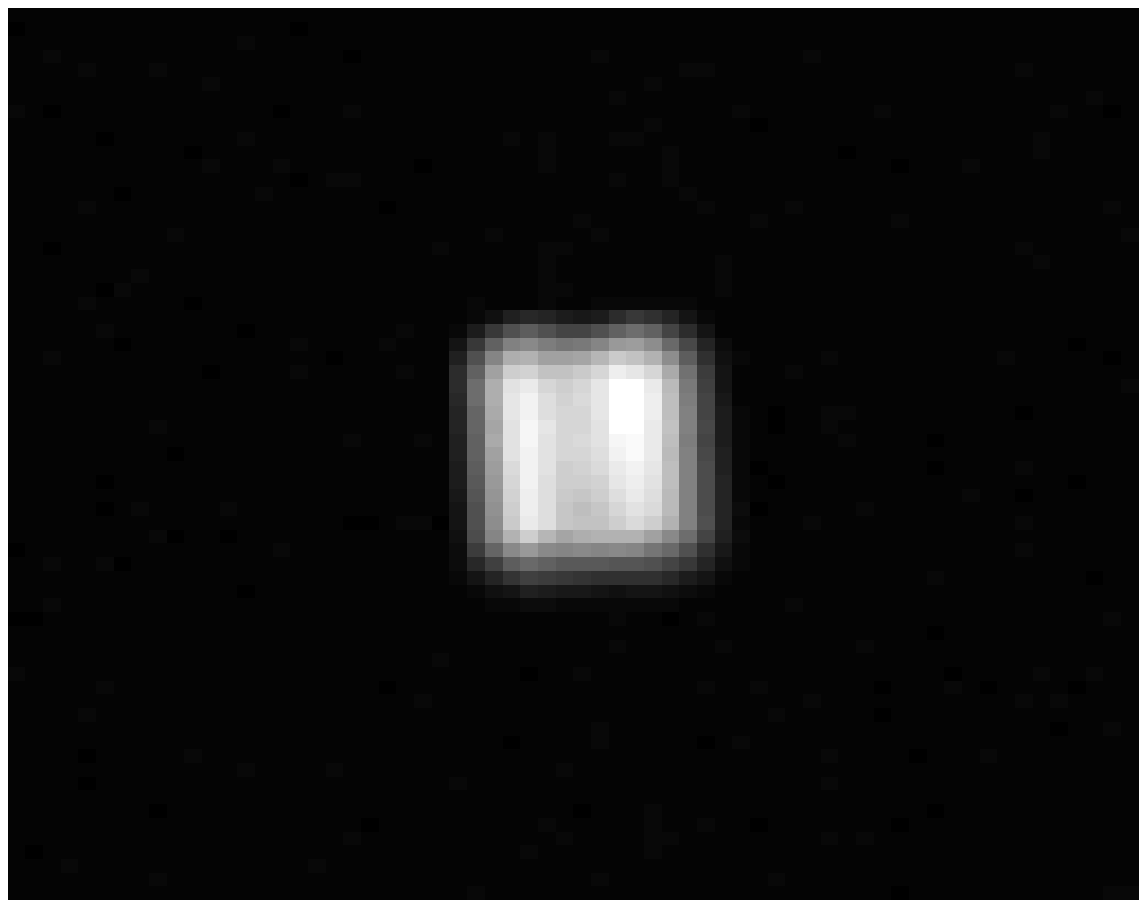}}
  \hspace{0in}
  \subfigure[]{\includegraphics[width=5cm]{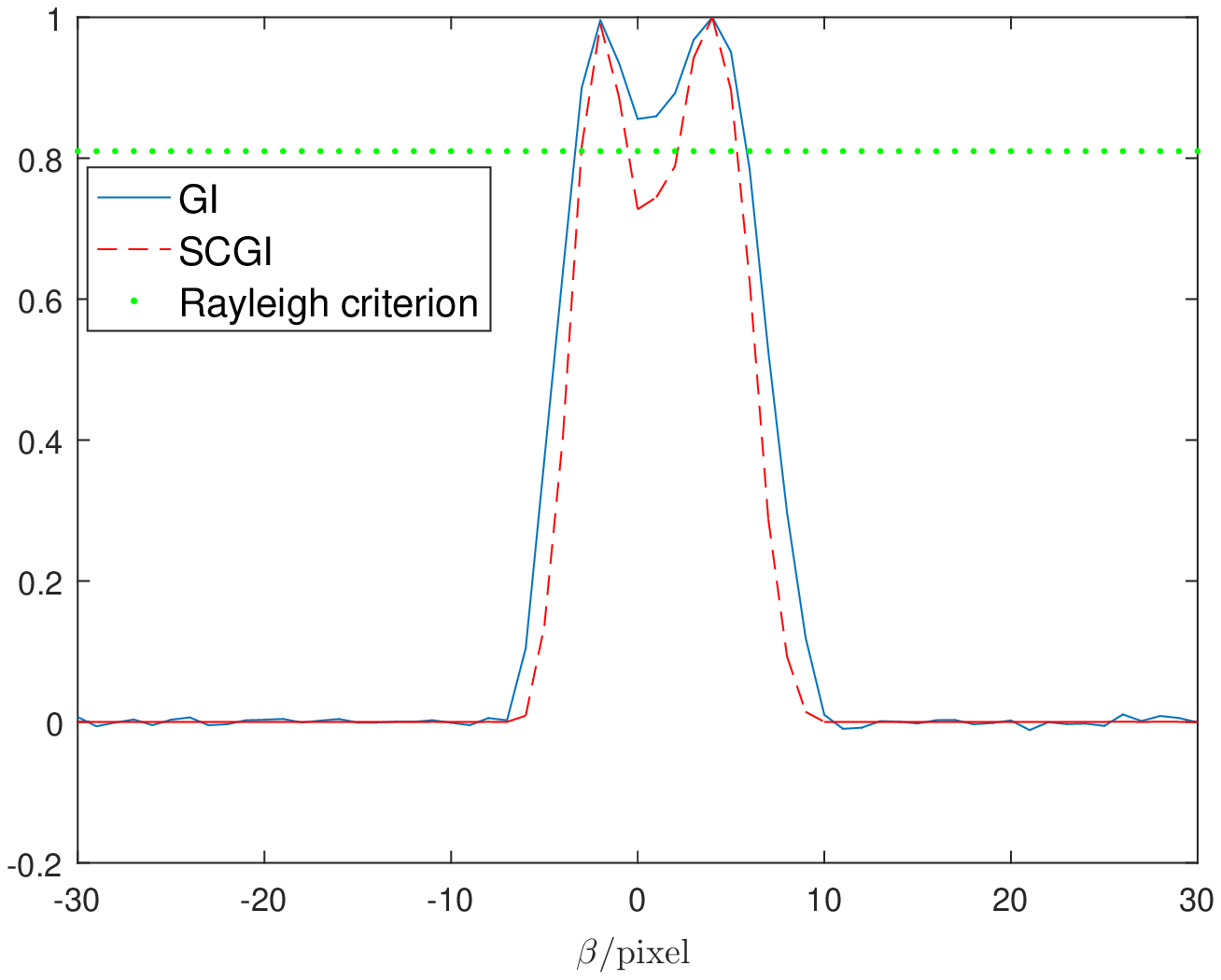}}
 \centering
  \caption{Results of experimental; (a),(b) the results of image reconstructed by SCGI and GI when $\Delta s_{o}=0.4m$, respectively; (c) the normalized horizontal section of the experimental results of (a) and (b); (d),(e) the results of image reconstructed by SCGI and GI when $\Delta s_{o}=0.85m$, respectively; (f) the normalized horizontal section of the experimental results of (d) and (e);}\label{4}
  \label{fig:false-color}
 \end{figure*}
Compared with GI, the resolution of SCGI is better, when $L(\beta) \neq 0$. The detailed verification is as follows: We use the Rayleigh criterion to describe the resolution of the GI\cite{paur2016achieving}. Rayleigh criterion specifies the minimum separation between two incoherent point source ($\alpha_{o}$ and $\alpha_{o}^{'}$. For simplification, we set $\alpha_{o}$=-$\alpha_{o}^{'}$) that may be resolved into distinct object\cite{paur2016achieving}. For the GI, because that the form of intensity PSF is $\text{sinc}^{2}$ function, so Rayleigh-distance set as $d_{1}=|\alpha_{o}-\alpha_{o}^{'}|$ when $\frac{\Delta G^{(2)}(I_{0}, 0)}{\Delta G^{(2)}(I_{0}, \alpha_{o})}\approx 0.81$. We can get the result is that $\frac{\kappa_{2}(0)}{\kappa_{2}(\alpha_{o})} = 0.6561<0.81$ when $|\alpha_{o}-\alpha_{o}^{'}|=d_{1}$ by Eq.~(\ref{(10)}), the results means SCGI can enhance the resolution limit of the conventional GI system.\par
We verify our theoretical results via experiment. Our experimental setup is a conventional computational ghost imaging system, we used a projector as the light source. The diameter of light source is $3.3mm$ and the mean wavelength is $\lambda =550nm$. A double slit is taken as an object to measure the resolution of imaging. The slit width is $a_{1}=2\times 10^{-3}m$, slit center distance is $b_{1}=3\times 10^{-3}m$ and slit height is $g_{1}=8\times 10^{-3}m$. For simplification, we get different resolution image by manipulating $\Delta s$ ($\Delta s =s_{r}-s_{o}$) instead of $b_{1}$\cite{zeng2017influence,gatti2008three,ferri2008longitudinal}. We can get that $d_{1} \approx 2.3mm$ when $s_{o}=0.35m$ and $\Delta s=0.4m$, which is obvious smaller than $b_{1}$. When $\Delta s=0.85m$, $d_{1}\approx 3.1mm$ which is larger than $b_{1}$. Thus, in the experiment we set $s_{o}=0.35m$, $\Delta s$ as $0.4m$ and $0.85m$ respectively. Fig.~\ref{4} shows the experimental results of images reconstructed by GI and SCGI for different $\Delta s$, respectively.\par
When $\Delta s=0.4m$, the result of SCGI is shown in  Fig.~\ref{4}(a) and the result of GI is shown in Fig.~\ref{4}(b). Fig.~\ref{4}(a) and Fig.~\ref{4}(b) are shown that the image of double slit by SCGI is more clear than that by GI. The normalized horizontal section of the experimental results of Fig.~\ref{4}(a) and Fig.~\ref{4}(b) are shown in Fig.~\ref{4}(c). The result of Fig.~\ref{4}(c) shows that $\frac{\kappa _{2}(0)}{\text{max}(\kappa _{2}(\beta))}<\frac{\Delta G^{(2)}(I_{0}, 0)}{\text{max}(\Delta G ^{(2)}(I_{0}, \beta))}$. When $\Delta s=0.85m$, Fig.~\ref{4}(d-f) are shown that the image of double slit by SCGI is more clear than that by GI when $b_{1}<d_{1}$. Above results suggest that SCGI can enhance the resolution limit of GI system. Our analysis is that $\kappa_{2}(\beta)$ has more information than $\Delta G^{(2)}(I_{0}, \beta)$.\par
In conclusion, $\Delta G^{(2)}(I_{0}, \beta)$ is replaced by $\kappa_{2}(\beta)$ to obtain the image of object in GI system. We  call this protocol as Second-order Cumulants ghost imaging (SCGI). Our theoretical analysis and experimental results show that the resolution limit of conventional GI can be enhanced by SCGI without changing the experimental setup of GI. $\kappa_{2}(\beta)$ is the fluctuation information of $\Delta G^{(2)}(I_{0}, \beta)$, so the resolution can be further enhanced by SCGI to the super-resolution schemes, such as compressive sensing technique, low-pass spatial filter scheme and so on. \par

This work is supported by the Science \& Technology Development Project of Jilin Province (No.YDZJ202101ZYTS030).\par
\section*{DATA AVAILABILITY}
The data that support the findings of this study are available from the corresponding author upon reasonable request.
\section*{REFERENCES}

\nocite{*}
\bibliography{aipsamp2}

\end{document}